\documentclass[12pt,preprint]{aastex}

\usepackage{emulateapj5}

\shortauthors{Toft et al.}
\shorttitle{DRGs in the UDF}

\begin{document}
\bibliographystyle{aa}

\title{Distant red galaxies in the Hubble Ultra Deep Field\altaffilmark{1} }

\author{S. Toft\altaffilmark{2}, P. van Dokkum\altaffilmark{2}, M. Franx\altaffilmark{3}, R. I. Thompson\altaffilmark{4}, G. D. Illingworth\altaffilmark{5}, R. J. Bouwens\altaffilmark{5},  M. Kriek\altaffilmark{3}}

\altaffiltext{1}
{Based on observations made with the NASA/ESA \emph {Hubble Space Telescope}, which is operated by the Association of Universities for Research in Astronomy, Inc, under NASA contract NAS5-26555}

\altaffiltext{2}
{Department of Astronomy, Yale University, New Haven, CT 06520-8101, email: toft@astro.yale.edu, dokkum@astro.yale.edu}

\altaffiltext{3}
{Leiden Observatory, Postbus 9513, 2300 RA Leiden, Netherlands, email: franx@strw.leidenuniv.nl, mariska@strw.leidenuniv.nl}
\altaffiltext{4}
{Steward Observatory, University of Arizona, 933 Cherry Avenue, Tucson, AZ 85721, email: thompson@cosmos.as.arizona.edu}
\altaffiltext{5}
{University of California at Santa Cruz, 1156 High Street, Santa Cruz, CA 95064, email: gdi@ucolick.org, bouwens@ucolick.org}

\begin{abstract}
We take advantage of the Hubble Ultra Deep Field (UDF) data to study the restframe optical and ultra violet (UV) morphologies of the novel population of Distant Red Galaxies (DRGs). Six galaxies with $J-K_s>2.3$ are found to $K_s=21.5$, five of which have photometric redshifts $z_{phot} \ga 2$, corresponding to a surface density of 0.9 arcmin$^{-2}$. The surface brightness distributions of the $z_{phot}\ga2$ galaxies  are better represented by exponential disks than  R$^{1/4}$-laws.
Two of the $z_{phot} \ga 2$ galaxies are extended, while three have compact morphologies.
The restframe optical  morphology of the $z_{phot}\ga2$ galaxies is quite different from the restframe UV morphology: all the galaxies have 
red central components which dominate in the NICMOS $H_{160}$-band images, and distinct off-center blue 
features which show up in (and often dominate) the ACS images.  
 The mean measured effective radius of the $z_{phot}\ga2$ galaxies is $\langle r_e \rangle =1.9\pm1.4$ kpc, similar (within the errors) to the mean size of LBGs at similar redshifts. 
All the DRGs are resolved in the ACS images, while four are resolved in the NICMOS images.
Two of the $z_{phot}\ga2$ galaxies are bright X-ray sources and hence host AGN.
One of these galaxies is resolved in the ACS and NICMOS images, which means the AGN does not dominate its restframe UV-optical SED, while the other is unresolved in the NICMOS images and hence could have an AGN dominated SED.
The diverse restframe optical and UV morphological properties of DRGs derived here suggest that they have complex stellar populations, consisting of both evolved populations that dominate the mass and the restframe optical light, and younger populations, which show up as patches of star formation in the  restframe UV light; in many ways resembling the properties of normal local galaxies.
This interpretation is supported by fits to the broadband SEDs, which for all five $z_{phot} \ga 2$ are best represented by models with extended star formation histories and substantial amounts of dust.  
\end{abstract}


\section{Introduction}
The Faint InfraRed ExtraGalactic Survey \cite[FIRES,][]{labbe2003} discovered a highly clustered \citep{daddi2003} population of Distant ($z_{phot}\ga2$) Red Galaxies (DRGs) by selecting galaxies with $J-K_s>2.3$ \citep{franx2003}. 
At similar restframe optical luminosities, DRGs typically have older stellar ages (1-3 Gyr), larger stellar masses ($\sim$10$^{11}M_{\sun}$), more dust ($A_V$=1-3 mag) and higher star formation rates (SFR$\sim$100$M_{\sun}yr^{-1}$) than the complementary population of high redshift galaxies selected using the Lyman break technique \citep{steidel1999,shapley2001,shapley2003,forster-schreiber2004,vandokkum2004}.
The observed properties of LBGs suggests that they are on average younger than the more evolved DRGs.

Morphological studies of Lyman Break Galaxies (LBGs) using the Hubble Space Telescope (HST) suggest that they are compact ($r_e \sim 2$kpc) and have very similar restframe UV and optical morphologies \citep{giavalisco1996, lowenthal1997,dickinson2000, ferguson2004, bouwens2004}.
Here we present the first space-based study of the restframe optical morphologies of DRGs.
We take advantage of the Hubble Ultra Deep Field (UDF) NICMOS and ACS imaging of a $2.4\times2.4$ arcmin$^2$ area of the sky in the Chandra Deep Field South (CDFS), to study restframe UV to optical morphologies of DRGs. The unprecedented depth\footnote{The 10$\sigma$ AB limits in the filters F435W, F606W, F775W, F850LP, F110W, F160W are  29.2, 30.0, 29.7, 28.7, 27.0, 27.0., \citep{beckwith2004,thompson2004}} of the data allow for detailed morphological studies, even of very low surface brightness components.     

In the following we assume a standard cosmology with $\Omega_m=0.3$, $\Omega_{\Lambda}=0.7$, and $h=0.7$ \citep{spergel2003}. We refer to the HST filters F435W, F606W, F775W, F850LP, F110W and  F160W as $B_{435}$, $V_{606}$, $i_{775}$, $z_{850}$, $J_{110}$ and $H_{160}$. Magnitudes are in the Vega system, unless otherwise noted.

\section{Selection}
In addition to the public UDF data \citep{beckwith2004,thompson2004} we use a photometric catalog consisting of ACS data (in the $B_{435}$, $V_{606}$, $i_{775}$ and $z_{850}$-bands) taken as part of the GOODS survey,
\null
\vbox{
\begin{center}
\leavevmode
\hbox{%
\epsfxsize=8.8cm
\epsffile{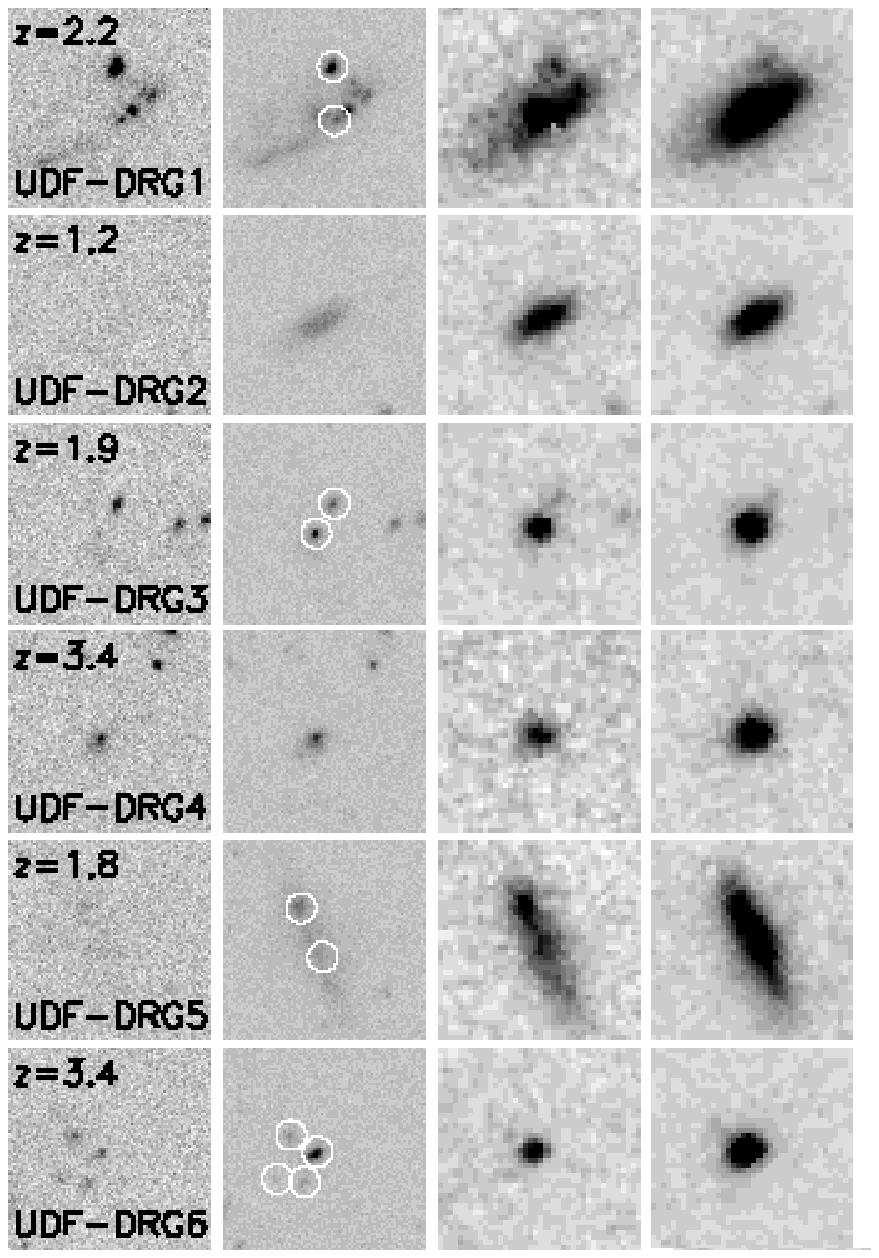}}
\figcaption{\small
Gallery of the six galaxies in the UDF with $K_s<21.5$ and $J-K_s>2.3$. The filters shown are (from left to right): $B_{435}+V_{606}$, $i_{775}+z_{850}$, $J_{110}$, $H_{160}$. The size of the stamps is $3\farcs6 \times 3\farcs6$. North is up, east is to the left. The images have been normalized such that a galaxy with  constant $f_{\lambda}$ would appear the same in all images.
The galaxies appear regular in the NICMOS images, and show a range of sizes.} 
\end{center}}
 combined with deep VLT/ISAAC data (in the $J$, $H$ and $K_s$-bands) taken as part of the ESO Imaging Survey (EIS) \citep{vanderwel2004}. In the following we refer to this catalog as the GOODS/EIS catalog.  
  
We use the GOODS/EIS catalog rather than the UDF catalog for selecting galaxies as it includes the same $J$ and $K_s$ filters that have been used previously \citep{franx2003}. 
The additional depth of the UDF $J_{110}$-band observation (relative to the ISAAC $J$-band observations) is less important as we are concentrating on relatively bright galaxies.
Galaxies were selected using the criteria:  $K_s<21.5$, $J-K_s>2.3$. Six galaxies in the UDF match these criteria.  A postage stamp gallery of the sample is shown in Fig. 1.

\section{Photometry and Photometric redshifts}
Photometric redshifts were derived both from $1\farcs8$ diameter aperture photometry derived from Point Spread Function (PSF)-matched ACS/NICMOS UDF images, and from $2 \arcsec$ diameter aperture photometry available in the GOODS/EIS catalog.
The GOODS/EIS dataset was found to provide stronger constraints on the redshifts than the UDF data set, due to the inclusion of the $K_s$-band and the fact that the ISAAC $J$-band is redder and narrower than the NICMOS $J_{110}$ band. 
Hyperz \citep{hyperz} was used to derive photometric redshifts. 
To limit the degrees of freedom in the fits, we used four empirical templates (E, Sbc, Scd and Im) to derive the best fitting photometric redshifts.  
In Tab 1, we list the best fitting photometric redshifts and other properties of the sample. 
 The median photometric redshift of the sample is $2.1\pm0.7$, in good agreement with the median photometric redshift found for DRGs in the HDFS \citep{franx2003}. One of the six galaxies in the sample has a photometric redshift significantly lower than 2.
The surface density of $z_{phot}\ga 2$ DRGs to $K_s=21.5$ is $\simeq0.9$ arcmin$^{-2}$, consistent with $\simeq1.0$ arcmin$^{-2}$ to $K_s=21.7$ found for a much larger sample in the MS1054 field \citep{forster-schreiber2004}.
Following \cite{franx2003} and \cite{forster-schreiber2004}, we next fixed the redshifts and fitted three \cite{bruzual2003} models (BC): a constant star formation model (CSF) with dust, a ``Tau'' model (exponentially declining star formation  with a characteristic timescale of $\tau=$1 Gyr) with dust, and a Single instantaneous burst model without dust.  
In Fig 2. we show SEDs of the six galaxies in the sample, along with the best fitting redshifted BC models (full black curves).
It is interesting to note that all six galaxies are better fitted by the ``Tau'' model with ongoing star formation and significant amounts of dust, than the single burst model with no dust (shown in dashed red curves).

\section{Morphologies}
The DRGs in our sample show a range of morphologies: in the NIR, three are extended and three are compact. Of the galaxies with $z_{phot}\ga 2$, three are compact and two are extended. 
The observed NIR (restframe optical) morphology of the $z_{phot}\ga2$ galaxies look quite different from the observed optical (restframe UV) morphology, highlighting the importance of deep NIR imaging of high redshift galaxies. 
In the NIR, all the $z_{phot}\ga2$ galaxies have red central components which dominate the NICMOS $H_{160}$-band images, while distinct blue off-center features show up in (and often dominate) the ACS images. We verified that these differences were not due to the different PSF and noise properties of the ACS and NICMOS images. 

The most pronounced differences between ACS and NICMOS are seen in two galaxies: UDF-DRG1, which has a luminous red bulge which dominates the light in the NIR bands and a more extended, patchy  blue disk which dominates the light in the optical bands; and UDF-DRG6, which in the NIR bands is a single compact source, but in the optical bands is composed of four close compact sources of comparable brightness. From the optical bands alone these two galaxies could easily have been classified as highly irregular galaxies, or galaxies undergoing major merging, while their NIR morphologies reveal that they are more likely relatively relaxed galaxies  with patches of ongoing star formation.

The extraordinary depth of the UDF data allow us to test whether the blue ``blobs'' seen in UDF-DRG 1, 3, 5 and 6 are due to substructure in the 
\null
\vbox{
\begin{center}
\leavevmode
\hbox{%
\epsfxsize=9.0cm
\epsffile{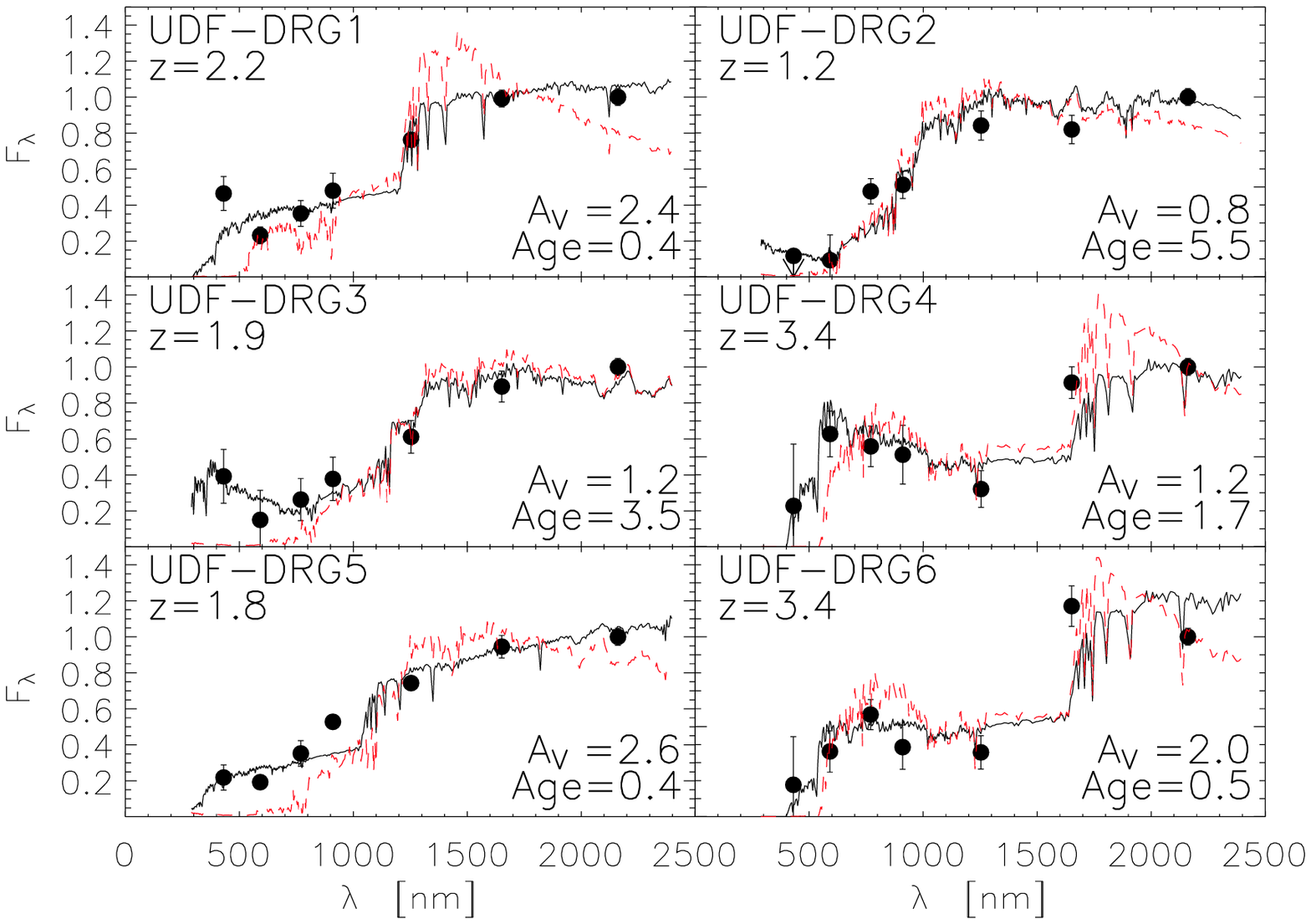}}
\figcaption{\small
SEDs ($f_{\lambda}$ versus $\lambda$) of the six galaxies in the sample. Solid circles are photometry from the GOODS/EIS catalog (normalized by $f_{\lambda}$ in  $Ks$-band). The full black curves are the best fitting (SB, Tau or CSF) BC models and the red dashed curves are the best fitting SB models (see text for details). The Tau model (with some dust extinction) provides the best fit to all six galaxies. The CSF models provide almost as good fits, but require more dust. The formal errors on the redshift from {\sc hyperz} are in the range $\sim$0.1-0.2, as are error estimates from monte carlo simulations.}
\end{center}}
galaxies, or chance alignment with unrelated sources at different redshifts. To do this we placed small (3 pix) apertures on the positions of the individual blobs 
(indicated on the $B_{435}+V_{606}$ images in Fig. 1) 
in the PSF-matched images and derived their SEDs and photometric redshift. These are all consistent with the blobs being substructure at the redshift of the main galaxies, although we note that the photometry in neighboring blobs are not fully independent.

\section{Sizes}
To better quantify the differences and similarities of the galaxies and their appearance at different wavelengths we modeled their surface brightness distributions  using {\sc galfit} \citep{peng2002}. We fitted a Sersic profile (convolved with the PSF, estimated from a calibration star): $I(r)=I(0)exp[b_n(r/r_e)^{1/n}]$  which has been shown to be a good representation of the surface brightness of a range of galaxy types from ellipticals ($n=4$) to disks ($n=1$).
Leaving the radial shape parameter $n$ and the effective radius $r_e$  free in the fitting resulted in a mean value of $n=1.4 \pm 0.9$.
Since there is some degeneracy between $n$ and $r_e$ in the fits, we next fixed $n$ to 1, to be able to compare the derived sizes of the galaxies with each other and with other high redshift galaxies.  
The resulting sizes (listed in Tab. 1) range from $0\farcs1$ to $0\farcs5$. At NIR wavelengths (in the $H_{160}$-band), the mean measured circularized radius\footnote{defined as $r_{e,c}=r_e\sqrt{(1+\epsilon)}$, where $r_e$ is the effective radius encompassing half the flux, measured along the semi-major axis, and $\epsilon=1-a/b$ is the ellipticity of light distribution} of the five $z_{phot} \ga 2$ galaxies is $\langle r_e \rangle_{NIR}=0\farcs23\pm0\farcs16$.
At the redshifts of the galaxies this corresponds to a mean physical size of $1.9\pm1.4$ kpc.   
At optical wavelengths most of the sizes are slightly larger, but less well defined due to the patchy light distribution. 
\null
\vbox{
\begin{center}
\leavevmode
\hbox{%
\epsfxsize=9.0cm
\epsffile{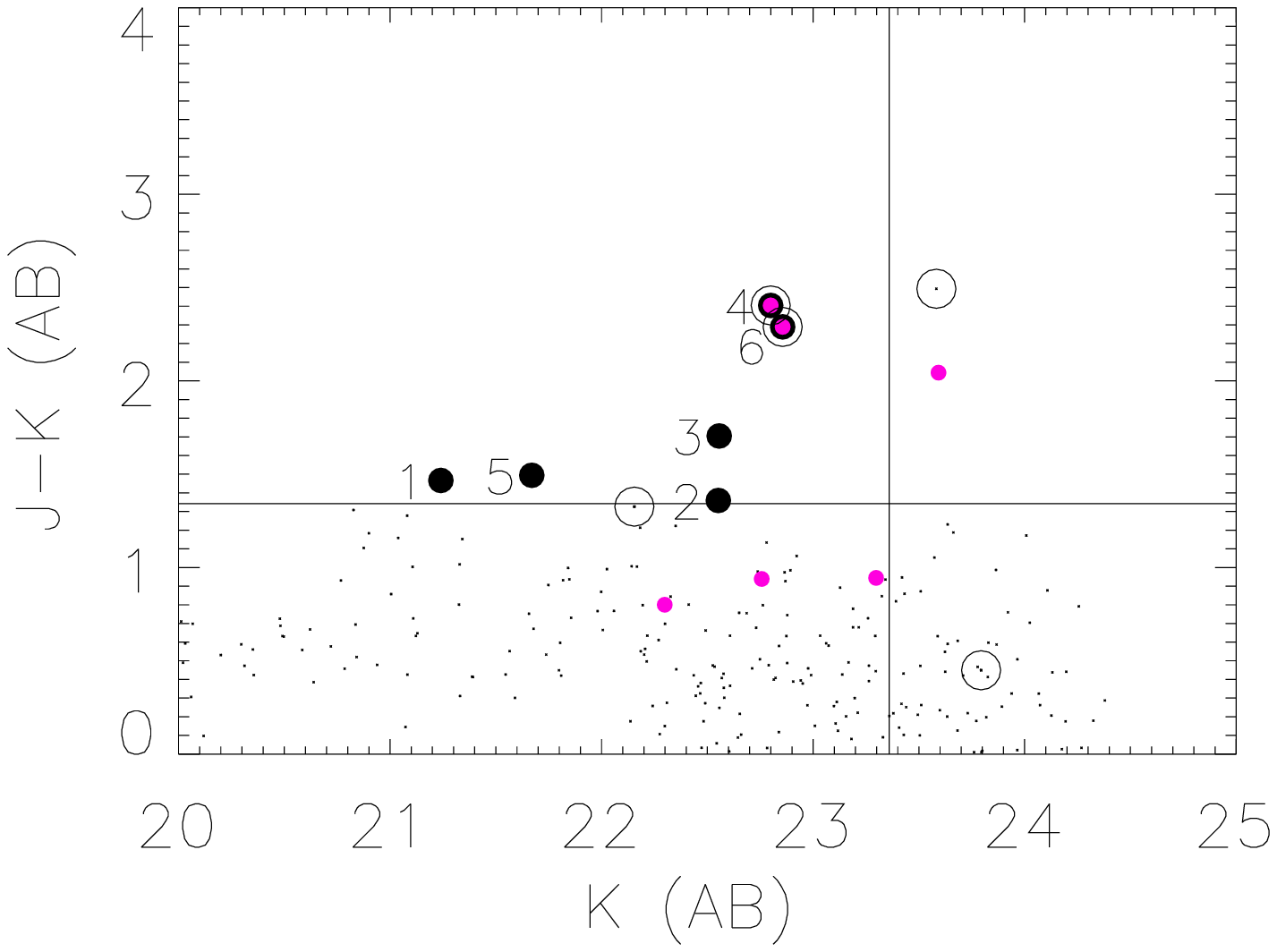}}
\figcaption{\small
Comparison of red galaxies selected here from their $J-Ks$ color (large solid symbols), from their $I_{775}-H_{160}$ color and $z_{phot}$ \citep[large open symbols,][]{chen2004}, and from their $z_{850}-3.6\micron (IRAC)$ color \cite[small filled (pink) symbols,][]{yan2004}. 
The colors and magnitudes of the  galaxies are from the GOODS/EIS catalog. Four of the \cite{chen2004} and four of the \cite{yan2004} galaxies are too faint to be detected in our $K_s$-band data. The three samples only have two galaxies in common (UDF-DRG4 and 6)}  
\end{center}}
All the DRGs are resolved in the ACS images. 
To determine whether the ``compact'' galaxies (UDF-DRG 3, 4 and 6) are resolved in the NICMOS data we repeated the size-fitting analysis on three stars (two stars in the field and a calibration star), to derive the mean size of  ``bonafide'' unresolved objects: $\langle r_e \rangle_{stars}=0\farcs06\pm 0\farcs03$. From this we conclude that UDF-DRG4 is marginally resolved, while UDF-DRG3 and 6 are consistent with being unresolved.  
Varying the fitting regions, and using different stars to model the PSF result in little variation in the derived $r_e$. 

The derived sizes can be compared to those of $z\sim3$ LBGs.
\cite{ferguson2004} find a mean size $\langle r_e\rangle \sim 2.3$ kpc, and \cite{bouwens2004} find $\langle r_e \rangle \sim 1.8$ kpc. As the luminosity range of the $z_{phot}\ga2$ galaxies in our sample ($M_V$ = $-21.4$ to $-23.3$) is comparable to that of the LBG samples we can compare the sizes directly \citep[see][]{trujillo2004}. The mean size for $z_{phot}\ga2$ DRGs derived here 
 is similar within the errors to the mean size of LBGs.

A possible caveat of this comparison is the fact that we are comparing DRG sizes derived in the restframe optical, with LBG sizes derived in the restframe UV. However, the difference between restframe UV and optical sizes derived here for our small sample of DRGs (see Tab.1) are much smaller than the galaxy-galaxy variation in sizes.

\section{AGN contribution}

Two of the DRGs in our sample are detected in the 1Ms Chandra Deep Field South X-ray catalog \citep{giacconi2002} to  $f_x> 10^{-16}$ ergs cm$^{-2}$s$^{-1}$. 
The [0.5--2.0] kev band X-ray fluxes of UDF-DRG 1  and 3 ($f_x=73$ and $81\times10^{-17}$ergs\,cm$^{-2}$s$^{-1}$ respectively) are comparable to
\begin{footnotesize}
\begin{center}
{ {\sc TABLE 1} \\
\sc Properties of distant red galaxies in the UDF} \\
\vspace{0.1cm}
\begin{tabular}{lllll}
\hline
\hline
\vspace{0.1cm}

Id$^a$ & $K_s$ & $J-K_s$ & $z_{phot}^{b}$ & 
 $r_{e,c}^{c}$ \\ 
\hline
 
UDF-DRG1 /41   & 19.38  &  2.42  &  2.2  &  $0\farcs35/0\farcs40$  \\
UDF-DRG2 /372  & 20.69  &  2.32  &  1.2  &   $0\farcs17/0\farcs19$ \\
UDF-DRG3 /290  & 20.70  &  2.66  &  1.9  &   $0\farcs09^d/0\farcs03$ \\
UDF-DRG4 /746  & 20.93  &  3.26  &  3.4  &   $0\farcs14/0\farcs12$ \\
UDF-DRG5 /899  & 19.81  &  2.45  &  1.8  &   $0\farcs49/0\farcs53$ \\
UDF-DRG6 /1207  & 21.00  &  3.24  & 3.4 &  $0\farcs08^d/0\farcs10$   \\
\hline
\end{tabular}
\end{center}
{\footnotesize
$^a$\, Id in this letter / Id in the public NICMOS UDF catalog (vers.2.0)\\
$^b$\, Photometric redshift derived from the GOODS/EIS catalog\\
$^c$\, Circularized effective radius derived in the $H_{160}$/$V_{606}$ filters. The typical error is $\sim0\farcs1$\\
$^d$\, Consistent with unresolved

}
\end{footnotesize}
\vspace{0.25cm}
 the fluxes of the two DRGs with AGNs in the MS1054-03 field \citep{rubin2004}. The corresponding X-ray luminosities are too bright to be supported by star formation and must thus be of AGN origin (the X-ray flux, due to star formation, expected from DRGs $f_x\simeq10^{-17}$ergs s$^{-1}$cm$^{-2}$ \citep{rubin2004} is below the detection limit).
The fact that two of the $z_{phot}\ga2$ galaxies certainly hosts AGN allow us to test the AGNs contribution to the restframe optical emission. UDF-DRG1 is clearly resolved in the NIR bands, so its red restframe optical color is probably due to evolved stars, rather than the AGN. UDF-DRG3 on the other hand is unresolved in the NIR bands, and could thus be dominated by the AGN, rather than stars, in its restframe optical emission.

\section{Discussion}    
Two other studies have recently identified red galaxies in the UDF. \cite{chen2004} select 9 galaxies with $I_{775}-H_{160}>2$ and $z_{phot}>2.5$, six of which are argued to be best described by evolved templates with no dust, and three by dusty starbursts. \cite{yan2004} combine the HST data with mid-infrared observations obtained with the Infra Red Array Camera (IRAC) onboard the Spitzer Space Telescope, to select 10 galaxies with $f_{\nu}(3.6\micron)/f_{\nu}(z_{850})>20$, and argue that their SEDs are well represented by a two component model consisting of an old evolved population and a secondary young population.
In Fig. 3 we plot  $J-K_s$ vs $K_s$ for the three samples of red galaxies in the UDF. 

The two reddest DRGs in our sample make it into both the \cite{chen2004} sample (Object 09024 and 01927 in their Table 1) and the \cite{yan2004} sample (object 8 and 9 in their Table 1). Of the remaining seven/eigth galaxies in the \cite{chen2004}/\cite{yan2004}) samples, four/four are too faint to be detected in our data, while three/four are either too blue or too faint to make our selection criterion.
We conclude that while the different selection methods show some overlap, they are not redundant.

We note, that while the photo-z's derived for the two galaxies present in all three samples  agree reasonably well ($\pm0.2$), the  model parameters are uncertain:  we find both galaxies to be best fit by a 1Gyr Tau model with dust; \cite{chen2004} find UDF-DRG4 to be best fit by a 0.3Gyr Tau model without dust, and UDF-DRG6 by a dusty CSF model; and \cite{yan2004} fits a two component model without dust to both galaxies.
This demonstrates that while photometric redshifts derived from fits to broadband SEDs are relatively robust, parameters of the fitted models are in general less well determined and should be interpretted with caution.

The most characteristic property of the restframe optical and UV morphologies of the DRGs studied here is their diversity. Some are extended, some are compact, some have very different restframe optical and UV morphologies, while some look similar. 
The conclusion that can be drawn is that most of the DRGs in the present sample have composite stellar populations, with central populations of red evolved stars dominating the restframe optical morphologies and younger more patchy distributed blue stars dominating the restframe UV morphologies, while LBGs have similar restframe optical and UV morphologies \citep{dickinson2000}, and thus may be uniformly younger systems with star forming regions so luminous that they dominate even the restframe optical light. These results are in agreement with those derived from spectra and broadband SEDs of DRGs \citep{forster-schreiber2004} and LBGs \citep{pettini2001,shapley2003}. 
There might, however, be some overlap in the properties of restframe optical and UV selected galaxies as a small fraction of restrame UV selected LBGs at  $z\simeq3$ and Bm/Bx galaxies at $z\simeq2$ have recently been shown to have red optical colors and ages similar to the DRGs \citep{steidel2004,shapley2004}.  
 
Support from the Danish Natural Science Research Council, and from NASA through HST grant GO-09803-10-A and LTSA NNG04GE12G is gratefully acknowledged.






\end{document}